# Low-Cost Multi-Gigahertz Test Systems Using CMOS FPGAs and PECL

D.C. Keezer, C. Gray, A. Majid, N. Taher
Georgia Institute of Technology
School of Electrical and Computer Engineering, Atlanta, GA

**Abstract**

*This paper describes two research projects that develop new low-cost techniques for testing devices with multiple high-speed (2 to 5 Gbps) signals. Each project uses commercially available components to keep costs low, yet achieves performance characteristics comparable to (and in some ways exceeding) more expensive ATE. A common CMOS FPGA-based logic core provides flexibility, adaptability, and communication with controlling computers while customized positive emitter-coupled logic (PECL) achieves multi-gigahertz data rates with about $\pm 25$ps timing accuracy.*

## 1. Introduction

Technology roadmaps clearly predict rapidly increasing clock and data rates into the foreseeable future. While the performance of future automated test equipment (ATE) will continue to increase as well, these improvements tend to be incremental, and lag that of leading edge components. Therefore, there is a need for test methods that do not rely upon rapid improvement in ATE, but rather can quickly adapt to new (usually higher-performance) test requirements.

A general concept called "test support processor" (TSP) was introduced in [1]. A TSP is a customized circuit which is added to an existing automated test system in order to enhance either the performance or to provide additional test functionality. Because the TSP is customized for specific applications, it can take advantage of newly-developed components.

By extending the TSP to operate without the aid of automated test equipment, a self-contained miniature tester can be constructed [2,3]. The customized miniature tester might not have the wide range of features provided by ATE. However it can be designed to provide just the specific test features needed for a particular application, and often at a lower cost than that of commercial ATE.

An FPGA-based digital logic core (DLC) is used to provide a stand-alone programmable tester as shown in **Figure 1**. The DLC produces several hundred moderate I/O speed (100-400 Mbps) signals. These are formatted and/or multiplexed using PECL devices to create sub-nanosecond bit periods and multi-gigabit-per-second signals. The internal DLC design is described in Section 2. An RF clock source (usually an external instrument) provides a low-jitter (picosecond) timing reference. This serves as both a master clock (for synchronous applications) and as a reference for all timing-critical signals. In some applications, the RF clock is also provided to either the DLC or to the device under test (DUT), or both. A personal computer communicates through a Universal Serial Bus (USB) with the DLC, and provides high-level control of the tests (which otherwise are synthesized in the DLC).

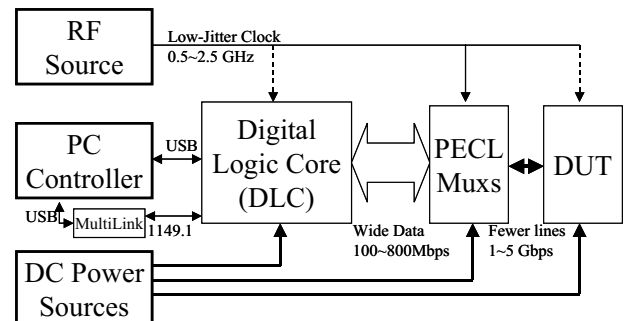

**Figure 1.** Using a programmable Digital Logic Core with high-speed PECL for testing a multi-gigahertz DUT.

In the first project, multi-gigahertz signals interface to optoelectronic components that modulate lasers of different wavelengths. The optical signals are combined at the transmitting end, and optically split at the receiving end (to recover the parallel data words). The present system is designed as a "test bed" for the evaluation of various OE and EO techniques, with nominal data rates of 2.5 Gbps per channel. Various signaling protocols are evaluated for the transmission of data packets through an optical switching network known as a "Data Vortex" [4,5]. For the test bed we create 5 high-speed data channels for both transmitting and receiving, to support a 4-bit parallel data word and source-synchronous clock. A lower-speed Framing bit, and four (lower-speed) data bits for routing address information are also generated. The end-application will require extending the word width to at least 64 bits, and increasing channel data rates to 10 Gbps at each wavelength, so that the aggregate data rate will be of the order of a Terabit-per-second. The objective is to provide low-latency transfer of small data packets within clusters of supercomputers. This application is further discussed in Section 3.

In the second project, a similar DLC is combined with multiplexing and sampling PECL circuits to create a self-contained tester for checking high-speed data paths in a wafer-probing environment. This "miniature tester" is designed to fit on the top side of a probe card, requiring only a source of power, a single RF clock signal, and a USB connection to a personal computer. The mini-tester produces a programmable data source up to 5 Gbps with 10ps timing resolution. A high-speed PECL sampling circuit is designed to capture the returned signal, also with 10ps resolution.

## 2. Digital Logic Core

The key features of the Digital Logic Core design are illustrated in **Fig. 2**. The central component is a 1-million gate FPGA (Xilinx XC2V1000), with over 200 I/O, each capable of running up to 800 Mbps. In addition, the DLC includes a specialized microcontroller chip for interfacing to a Universal Serial Bus (USB). Supporting these are a 12 MHz crystal oscillator for USB communications, and a FLASH memory to store the FPGA programming information. The FLASH is programmed from a personal computer through an IEEE1149.1 (boundary scan) interface. Once programmed, it loads the personalization data to the FPGA upon power-up. The program can be changed by overwriting the FLASH. This feature is very useful for quickly adapting the DLC to handle new test applications, or to make corrections in an existing design. State machines encoded in the FPGA, together with higher-speed PECL multiplexers and sampling circuits synthesize the desired tests in real time. A high-speed port to optional SRAM is also part of the design, although not used in the applications presented here. The SRAM can provide extended test pattern storage when algorithmic pattern generation is not feasible. About 200 signals are available from the FPGA that serve as general-purpose I/O to support specific test applications. In principle, these are capable of running at 800 Mbps, although we typically limit them to 300 or 400 Mbps in order to maintain sufficient of design margin. In some applications, these signals can serve directly as I/O for testing the DUT. However, higher (multi-gigahertz) speeds are obtained using additional PECL multiplexers, timing generators, and sampling circuits (see Sections 3 and 4).

## 3. Optical Test Bed Application

For the first project, we needed to create very precisely-timed logic signals to emulate the behavior of a parallel slice from a microprocessor-to-memory communication channel. These signals were used to control laser drivers which converted the signals to light pulses of differing wavelengths. These optical signals are then used to exercise and test a Data Vortex [4,5], an experimental switching fabric designed to address the issues associated with interfacing an optical packet interconnection network to high-performance computing systems. The Data Vortex can be used to interface processors and shared memory, which the DLC is emulating, through an ultrahigh capacity low latency communications system as illustrated in **Fig.3**.

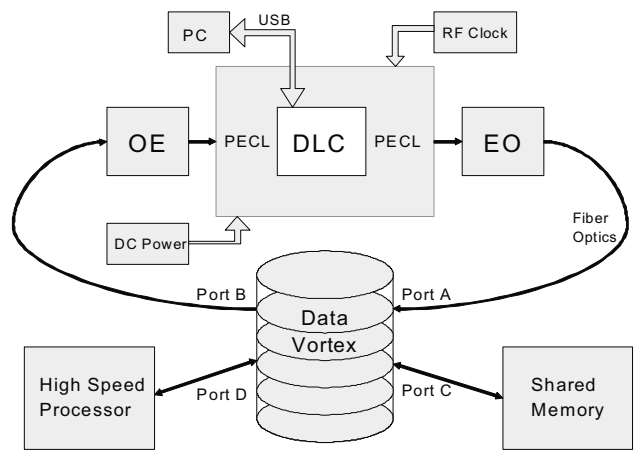

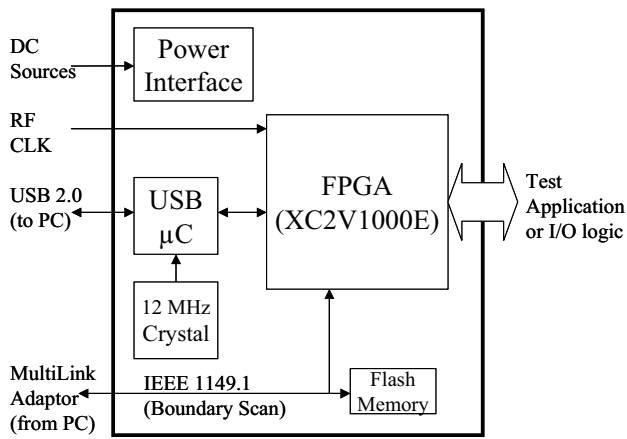

**Figure 2.** Digital Logic Core used for several test applications.

**Figure 3.** Optoelectronic test application (Data Vortex).

An example of the desired wave shapes is shown in **Fig.4**. Here four example data signals (each 32 bits in length) are synchronously produced to emulate part of a much wider data bus. These are precisely aligned in time with a source-synchronous reference clock. A much slower "Frame" bit is also produced to signal when the data is valid. Four "Header" channels carry the routing address data which is used within the Data Vortex to optically route the message to the desired port.

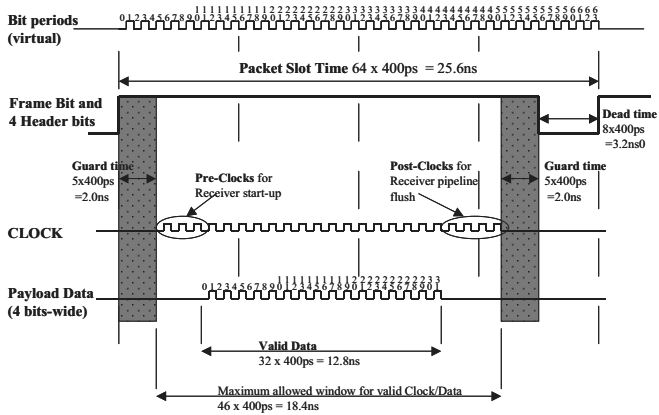

**Figure 4.** Test stimuli signals needed for the Optical Test Bed application.

The relative timing for leading and trailing edges for both data and Framing/Header signals must be controlled with 10ps resolution in the Optical Test Bed. A 10ns range for the placement of these edges is also required. The design of the PECL circuits to produce these signals paid careful attention to maintaining timing accuracy and to minimizing jitter.

**Figure 5** shows an early version of the Optical Test Bed electronics. This includes the DLC as well as PECL circuits for the transmitting and receiving functions. The SMA connectors on the right side of the board are for connection to the optical components.

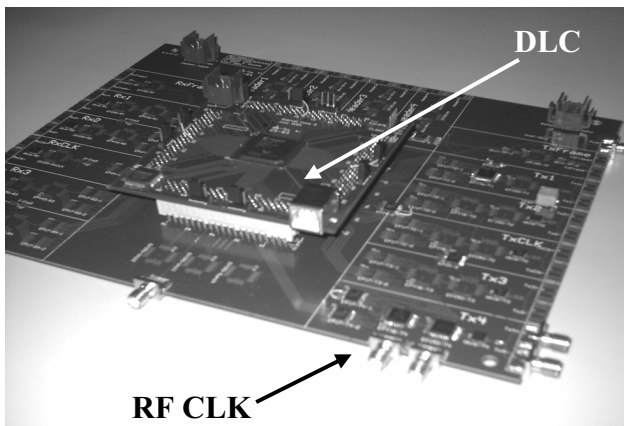

**Figure 5.** Transmitter and Receiver electronics used for the Optical Test Bed application.

Some representative waveforms are shown in **Figure 6.** Four data words are being controlled by the DLC and serialized by the PECL circuitry at 2.5 Gbps data rate.

We independently measured the 20 to 80 percent rise and fall times and found them to be in the range of 70 to 75ps. These fast transition times were produced using silicon germanium (SiGe) buffers in the final output stage.

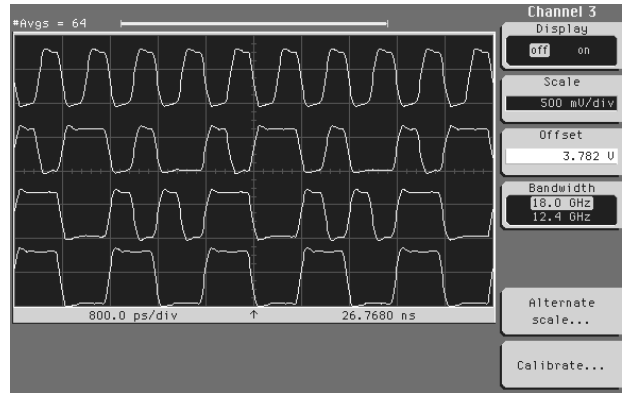

**Figure 6.** Example 2.5 Gbps transmitter data signals for the Optical Test Bed application. Note that 2.5 Gbps is the target rate for this application.

**Figure 7** shows an eye diagram with the output operating at the project target rate of 2.5 Gbps. For this test, the output waveform is a pseudo-random bit pattern produced by an LFSR in the DLC. In addition to fast rise and fall times, the SiGe output buffers introduce very little jitter, which is measured at the crossover point. For this signal, jitter was measured to be 46.7ps peak-to-peak, resulting in a usable eye opening of 0.88 unit intervals (UI).

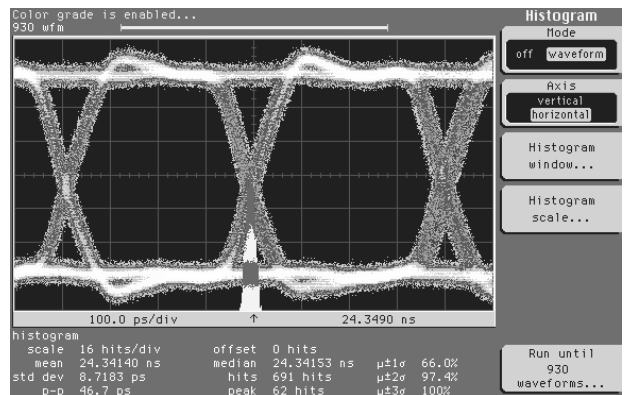

**Figure 7.** Example 2.5 Gbps eye diagram.

A similar measurement is shown in **Figure 8** operating at a data rate of 4.0 Gbps (considerably higher than the target rate for the project). The measured jitter at the crossover point was 47.2ps p-p with a usable eye opening of 0.81 UI and no visible signal attenuation. This bit rate is at the upper limit of some of the individual PECL components.

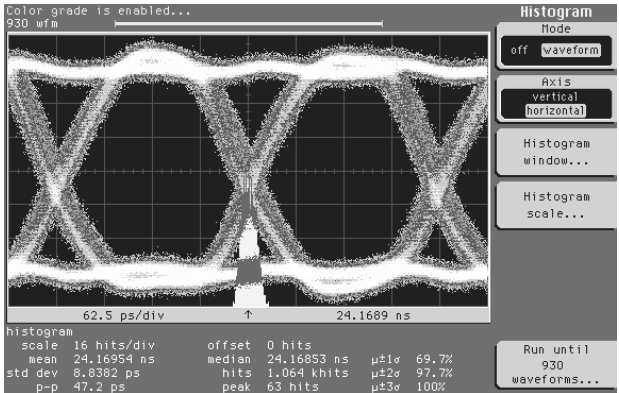

**Figure 8.** Example 4.0 Gbps eye diagram.

Shown in **Figure 9** is a measurement of a single falling edge of an output data channel, exhibiting only 24ps peak-to-peak jitter, and about 3.2ps rms. Unlike the previous eye diagrams, this jitter measurement does not include data dependent effects and therefore is only related to random jitter in the internal clock and the logic circuitry.

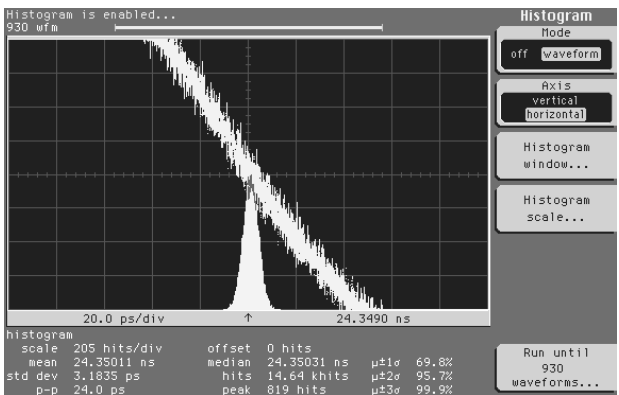

**Figure 9.** Jitter measurement for a single transition edge (24ps p-p).

**Figure 10** shows the controllability of the output voltage levels. The high logic level is shown at its maximum value and at three lower values in 100mV steps. Similar control is available on the low logic level and the midpoint bias, as illustrated in **Figure 11**. By controlling these values, a wide range of amplitude swings and midpoint bias values can be generated for characterizing the Data Vortex performance under non-ideal signal conditions.

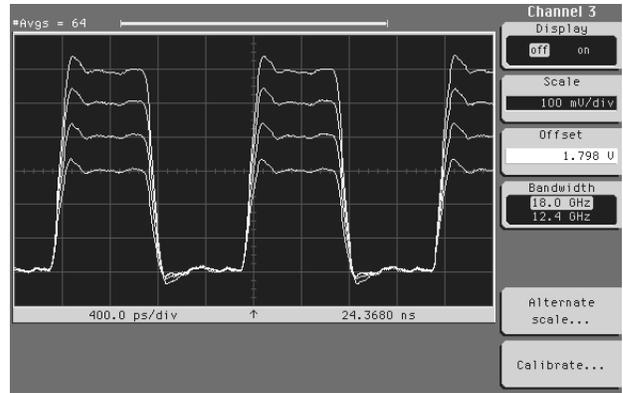

**Figure 10.** Adjusting the high logic level in 100mV steps. This example signal is running at 1.25 Gbps.

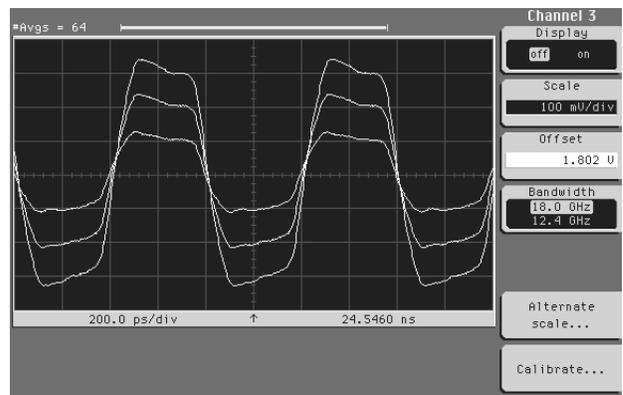

**Figure 11.** Adjusting the logic amplitude swing in 200mV steps. This example signal is running at 2.5 Gbps.

## 4. Miniature tester for wafer-level probing

The Miniature Tester may be used to test Wafer Level Packaged (WLP) devices if they have miniature compliant leads. These leads are fabricated on the wafer surface as a means to interconnect to the next packaging level when the chips are mounted. The extremely high density of the WLP devices makes connecting to the interface difficult. Therefore, as illustrated in **Figure 12**, an interposer is used to redistribute the high density WLP signals to a macroscopic scale (similar to a micro-BGA). A customized "Mini-Tester" based on the Digital Logic Core (DLC) is illustrated as a self-contained module mounted to the top side of a multi-layer printed circuit board which serves in place of the traditional probe card. Connections to the miniature tester are limited to: DC power, USB, and a high-performance (low-jitter, multi-gigahertz) clock input.

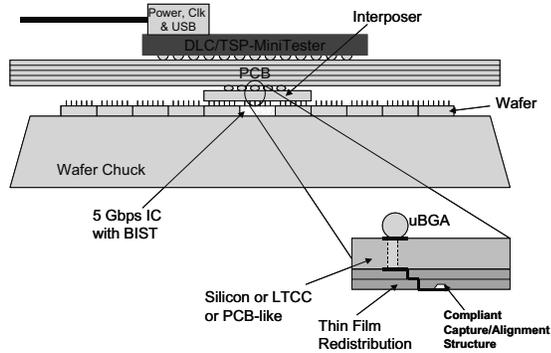

**Figure 12.** High-speed wafer-probe testing of wafer-level packaged (WLP) devices using a "miniature tester" and a high-density interposer.

When WLP compliant leads are available on all die sites, the miniature tester may be replicated in array form as illustrated in **Figure 13**. Functional testing can then be done in parallel, increasing production throughput by an order of magnitude. The complexity of the PCB is minimized by using only a small number of signals for each mini-tester, taking advantage of BIST features of the DUT. This strategy is a logical extension of existing parallel tests (such as used in memory testing) and is an extension of the TSP/mini-tester approach employing highly aggressive WLP testing.

An example of the miniature tester prototype board can be seen in **Figure 14**. The DLC is used again to provide general purpose communication to a PC through the USB, and to implement the state-machine logic for controlling some specific tests. The tests are designed to demonstrate high-speed (~5Gbps) signal propagation through the compliant lead structures. Therefore both stimulus generation and signal sampling is required at this high rate. This is accomplished by adding PECL circuits to the DLC which implements signal-multiplexing and picosecond sampling as shown in **Figure 15**.

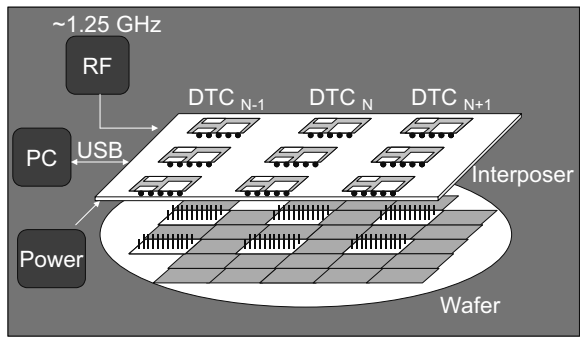

**Figure 13.** Parallel high-speed wafer probing using multiple miniature testers.

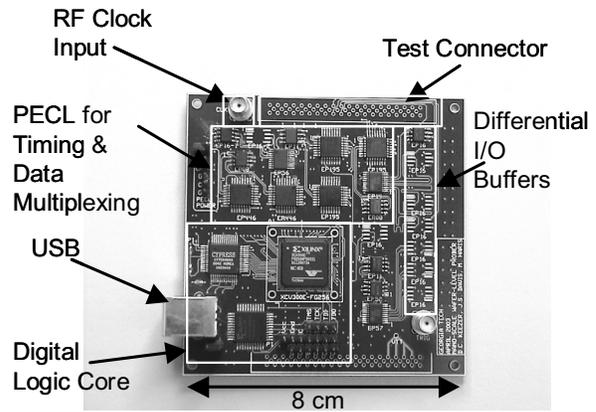

**Figure 14.** Prototype miniature tester with embedded Digital Logic Core.

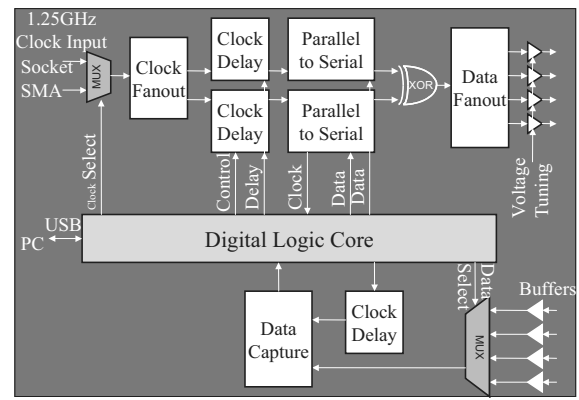

**Figure 15.** PECL logic used in a miniature tester with embedded Digital Logic Core.

Since the CMOS I/O in the DLC is limited to about 300-400 Mbps per signal, two groups of eight such signals are multiplexed to form two independent data sources at higher speeds (up to 2.5 Gbps). These are then combined in a second-stage multiplexer to obtain double the final signal (up to 5.0 Gbps). **Figure 16** shows an eye diagram at 1.0 Gbps. This signal shows a wide eye opening, sharp transitions, and peak-to-peak jitter of about 50ps (measured separately). At this frequency, we have an eye opening of about 0.95 UI.

As shown in **Figure 17**, the eye opening at 2.5 Gbps is slightly smaller, about 0.87UI. **Figure 18** shows bit patterns generated at the target rate of 5.0 Gbps. At such high speeds the rise time of the I/O buffers, measured at 120ps for 20% to 80%, begins to limit amplitude swing. Nevertheless, the 5.0 Gbps data eye diagram shown in **Figure 19**, still shows open eyes. At this speed the low jitter (~50ps) is proportionately larger, decreasing the eye opening to about 0.75UI.

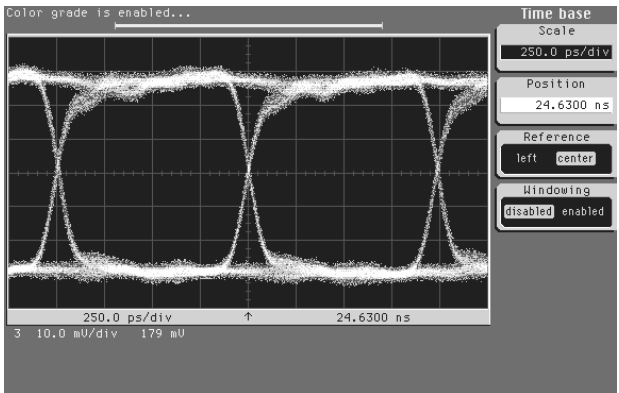

**Figure 16.** Measured 1.0 Gbps eye diagram, produced by the miniature WLP tester.

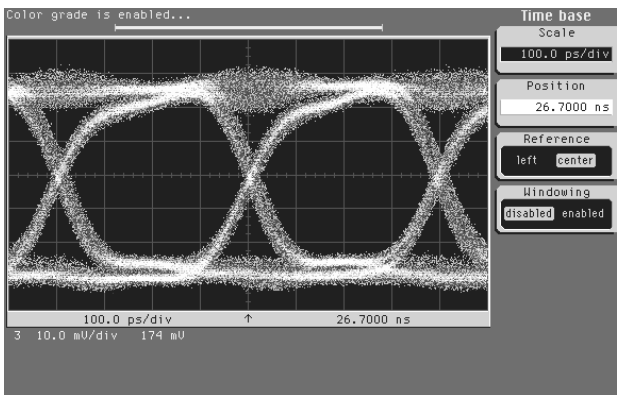

**Figure 17.** 2.5 Gbps eye diagram.

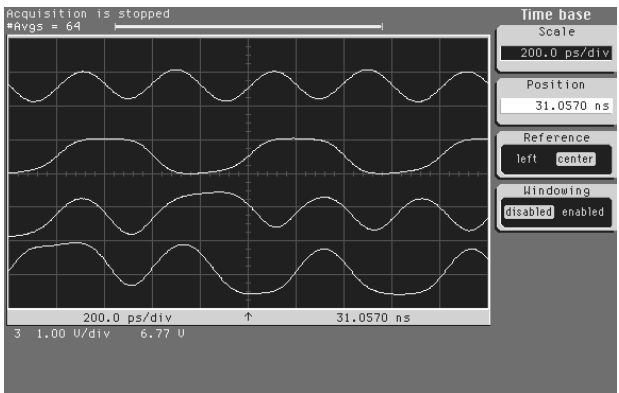

**Figure 18.** Example 5.0 Gbps signals (Target rate for this application).

## 5. Summary and Conclusions

In this paper we have demonstrated two examples of how a CMOS FPGA-based DLC can be used as the central controlling logic for multi-gigahertz test applications. The DLC was used to handle hundreds of signals each running in the few hundred Mbps range, and PECL was used to multiplex these signals to speeds as fast as 5 Gbps.

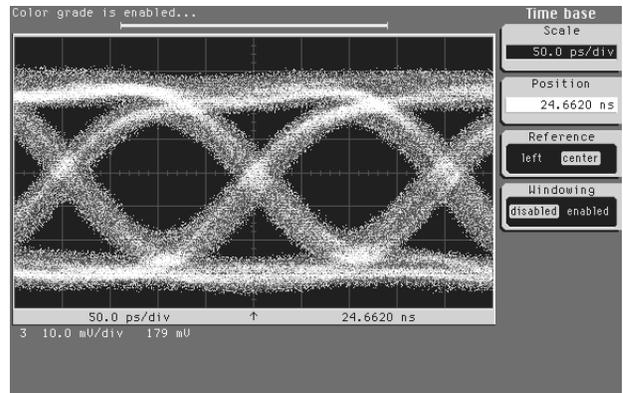

**Figure 19.** 5.0 Gbps eye diagram, produced with the miniature WLP tester.

The 200ps bit period at 5 Gbps means that accuracy of signal timing is critical. We have demonstrated timing accuracy control to about $\pm 25$ps, resulting in data eye openings of about 0.75 UI at 5.0 Gbps. Therefore the use of low-cost commercial off-the-shelf components results in test systems that are significantly lower in cost than conventional ATE.

### Acknowledgments

This work was supported in part by the National Science Foundation under contract EEC-9402723. The National Science Foundation supports the Georgia Tech Packaging Research Center (PRC) as a NSF Engineering Research Center.